\documentclass[aps,prb,twocolumn,floatfix,showpacs,amsmath,amssymb]{revtex4}
\usepackage{amssymb}
\usepackage{amsmath}

\usepackage{graphicx}
\usepackage[hypertex]{hyperref}
\begin{document}

\title{High-pressure phases of a hydrogen-rich compound: tetramethylgermane}

\author{Zhen-Xing Qin,$^{1}$ Chao Zhang,$^{2}$ Ling-Yun Tang,$^{1}$ Guo-Hua Zhong,$^{3}$ Hai-Qing Lin,$^{2,3}$ and Xiao-Jia Chen$^{4,1,5}$}
\affiliation{$^{1}$Department of Physics, South China University of Technology, Guangzhou 510640, China\\
$^{2}$Beijing Computational Science Research Center, Beijing 100084, China\\
$^{3}$Shenzhen Institutes of Advanced Technology, Chinese Academy of Sciences, Shenzhen 518055, China\\
$^{4}$Geophysical Laboratory, Carnegie Institution of Washington, Washington, DC 20015, USA\\
$^{5}$Center for High Pressure Science and Technology Advanced Research, Shanghai 201203, China}

\date{\today}

\begin{abstract}
The vibrational and structural properties of a hydrogen-rich Group IVa hydride, Ge(CH$_3$)$_4$, are studied by combining Raman spectroscopy and synchrotron X-ray diffraction measurements at room temperature and at pressures up to 30.2 GPa. Both techniques allow the obtaining of complementary information on the high-pressure behaviors and yield consistent phase transitions at 1.4 GPa for the liquid to solid and 3.0, 5.4, and 20.3 GPa for the solid to solid. The four high-pressure solid phases are identified to have the cubic, orthorhombic, monoclinic and monoclinic crystal structures with space groups of \emph{Pa}-3 for phase I, \emph{Pnma} for phase II, \emph{P}2$_1$/\emph{c} for phase III, and \emph{P}2$_1$ for phase IV, respectively. These transitions are suggested to result from the changes in the inter- and intra-molecular bonding of this compound. The softening of some Raman modes on CH$_3$ groups and their sudden disappearance indicate that Ge(CH$_3$)$_4$ might be an ideal compound to realize metallization and even high-temperature superconductivity at modest static pressure for laboratory capability.
\end{abstract}
\pacs{78.30.Ly; 64.70.K-; 64.30.-t}

\maketitle

\section{INTRODUCTION}

The tetra-alkyl compounds of group IVa elements have attracted much attention from the scientific community due to their highly symmetrical character. The skeletal vibrations of their molecules have been investigated using Raman and Infrared spectra since 1930s [\onlinecite{Young1947,shimizu1960,Overend1960,Watari1978,Kohlrausch1932,Rank1933,Rank1935,Anderson1936,Edsall1937,Sheline1950,Silver1940,Perry1983}].
In the molecular structure of these compounds, the group IVa element is tetrahedrally coordinated by methyl groups (CH$_3$), making the molecule with the $T$$_d$ symmetry. The CH$_3$ groups are expected to play an important role in understanding the interesting physical and chemical properties of the tetra-alkyl compounds of group IVa
elements.\cite{Mones1952,Wolf2010,Valerga1970,Krebs1989,Fleischer2003,Warmuth1984,Prager1983}
At low temperature, the CH$_{3}$ groups become non-equivalent and exhibit intermolecular interactions. \cite{Warmuth1984,Prager1983} In addition, the CH$_3$ groups show various interesting behaviors at high pressures. Upon compression, the rotation of CH$_3$ groups have been restricted in some CH$_3$-rich compounds, such as CH$_3$HgM (M=Cl, Br, I)\cite{Adams1988} and (CH$_3$)$_2$XM (X=Sn or Tl).\cite{Rush1966,Adams1993} The CH$_3$ groups display different rotational angles in cubic Si(CH$_3$)$_4$ (TMS) at 0.58 GPa [\onlinecite{Gajda2008}]. Therefore, understanding the behavior of CH$_3$ groups in the tetra-alkyl compounds of group IVa elements, especially the variance of CH$_3$ group under pressure, is important for condensed matter physics, materials science, and chemistry.

Group IVa hydrides also provide an alternative way to metallic hydrogen which was predicted to be a superconductor with high transition temperature in monatomic and molecular phases. In group IVa hydrides, the hydrogen atoms probably have undertaken chemical precompression by the group IVa atoms within the unit cell,\cite{Ashcroft2004} and thus the chemical pressure environments in these hydrides may greatly reduce the physical pressure necessary for metallic hydrogen. Several experimental and theoretical efforts are currently underway to examine this prediction, such as
SiH$_4$ [\onlinecite{Feng2006,Pickard2006,Yao2007,Degtyareva2007,chen2008,chenxj2008,Eremets2008,Kim2008,Martinez2009,Degtyareva2009,Yan2010}],
GeH$_4$ [\onlinecite{Martinez2006,Li2007,Gao2008,Zhang2010,zhang2010}], SnH$_4$ [\onlinecite{Tse2007,Tsejs2007,Gonzalez2010,Gao2010}], and PbH$_4$ [\onlinecite{Zaleski2011}]. However, very recently experiment shows the possible decomposition of SiH$_4$ under irradiation from X-ray and lasers.\cite{Hanfland2011,Strobel2011} Excitingly, Si(CH$_3$)$_4$, one of the tetra-alkyl hydrides of group IVa element, was found to be stable up to 140 GPa in our recent work,\cite{qin2011} although it remains insulating. Above 96 GPa, the sudden disappearing of original vibrational modes and appearing of new Raman modes make the metallization of tetramethylsilane more complex. In addition, it is suggested that the homologous hydrides with heavier group IVa atom would yield lower metallization pressure, due to the weaker chemical bonds which can be dissociated at high pressures.\cite{Tsejs2007} Therefore, the investigation of heavier group IVa hydrides is in great demand.

Tetramethylgermane (TMGe), Ge(CH$_3$)$_4$, as one of heavier group IVa hydrides, belongs to a class of non-polar molecular compounds [Fig.~1(a)]. At low temperature, only one motification of TMGe was observed in the temperature range 15-300 K [\onlinecite{Valerga1970}]. It was found that the entropies of the potential barrier to rotation of the CH$_3$ groups of TMGe is surprisingly low when considering the trends of the potential barriers in other methyl compounds of the Group IV elements. Although, the crystal structures of TMGe were predicted by global lattice-energy minimizations using force-field methods,\cite{Wolf2010} no high-pressure phases have been determined experimentally. Therefore, it is of paramount important to investigate the phase transitions and stability of TMGe under pressure, especially the inter- and intra-molecular interactions of CH$_3$ gourps.

\begin{figure}[tbp]
\vspace{0cm}
\includegraphics[width=\columnwidth]{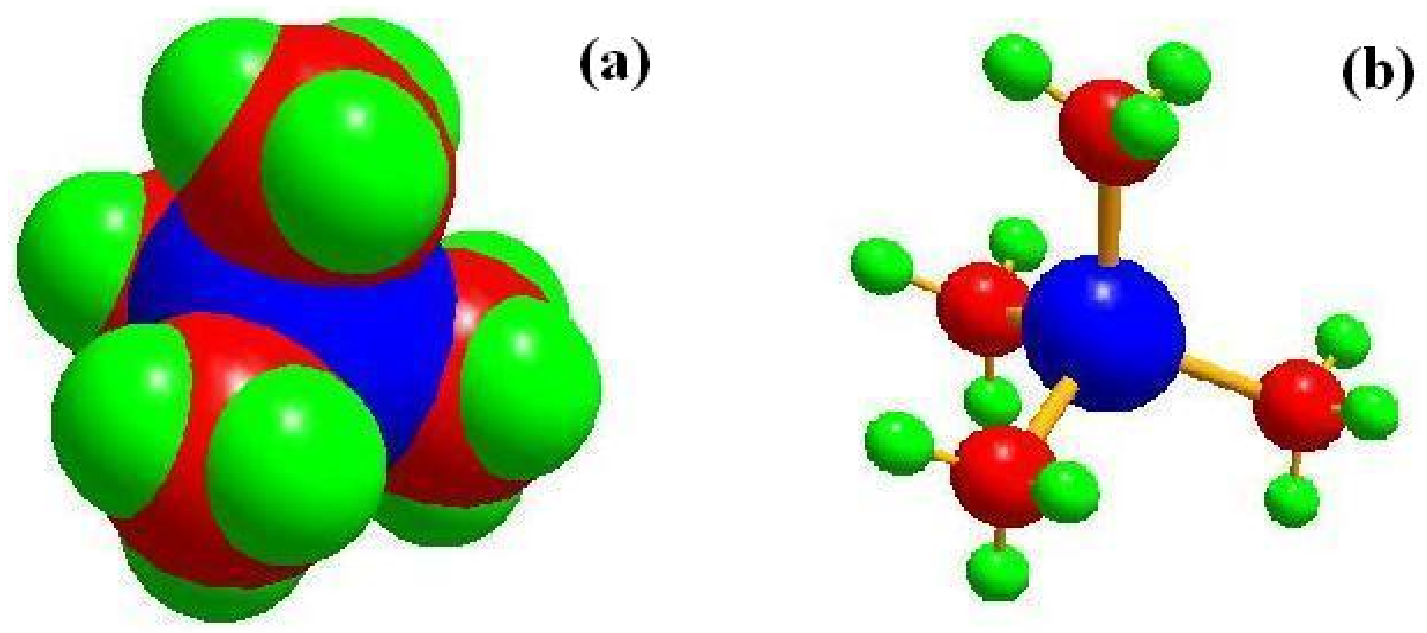}
\vspace{-0.5cm}
 \label{model} \caption{ \label{model} (Color online)
The configurations of TMGe with respect to ideally tetrahedral \emph{T}$_\emph{d}$ (-43\emph{m}) symmetry. (a) Cup model of TMGe shown to illustrate almost spherically shaped molecules in a close-packed stacking and (b) ball-and-stick model of TMGe manifested that one bond distance C-H, one Ge-C,one Ge-C-H/C-Ge-C angle could describe the molecular geometry.}
\vspace{-0.2cm}
\end{figure}

In the present work, the high-pressure behaviors of TMGe are investigated by combining Raman scattering and synchrotron X-ray diffraction (XRD) techniques up to 30 GPa using the diamond anvil cells (DAC). Several possible phase transitions are identified at 1.4, 3.0, 5.4, and 20.3 GPa by Raman spectroscopy, and their structures are also determined based on the the obtained synchrotron XRD data. The variation of CH$_3$ groups with pressure is examined over the whole pressure range studied. The structural and vibrational features are provided and discussed for this hydrogen-bearing compound.

\section{EXPERIMENTAL DETAILS}

TMGe (m.p. 185 K, b.p. 316 K) as transparency liquid with 98\% purity was purchased from Alfa-Aesar and used without further purification. The high-pressure experiments for TMGe were carried out using DAC with the culets of 300 microns. A hole of~$\sim$100 microns in diameter drilled in a preindented tungsten gasket served as the sample chamber. To avoid volatilizing, the bottom of DAC was put into ice-water mixture half an hour before loading the sample. Liquid TMGe was loaded into the chamber of DAC with a syringe. Because of liquid sample, no pressure medium was used and ruby grains had been placed previously for calibrating pressure. Raman spectra were measured in a backscattering geometry with a spectrometer (with 1200 mm$^{-1}$ grating) equipped with a di-monochromator and a charge coupled device detector, giving a resolution of ~1-2 cm$^{-1}$. Radiation of 633 nm from a solid-state laser (50 mW CW) was used for the excitation of the Raman spectra and all spectra were measured at room temperature.

\begin{figure}[tbp]
 \vspace{-0.5cm}
\includegraphics[width=\columnwidth]{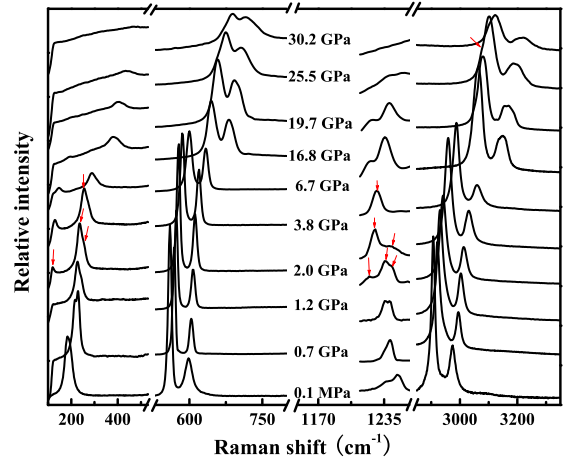}
 \vspace{-1cm}
\label{raman} \caption{(Color online) Representative Raman spectra of TMGe in the full spectral regions at ambient conditions upon compression to 30.2 GPa. The red arrows indicate the sudden changes of Raman modes. } 
\vspace{-0.2cm}
\end{figure}

The same DAC was employed for high-pressure synchrotron XRD experiment. Considering the volatilizing of samples, pressure was increased to 0.6 GPa after the sample was loaded. Synchrotron XRD data were collected at the X17C beamline at the National Synchrotron Light Source of Brookhaven National Laboratory via
angle-dispersive diffraction techniques using monochromatic radiation $\lambda=0.4067~\AA$. The sample-to-detector distance and the image plate orientation angles were calibrated using CeO$_2$ standard. The two-dimensional diffraction images were converted to 2$\theta$ versus intensity data plots using the FIT2D software.

In both measurements, the data shown here were collected in the compression cycle. However, we also performed the measurements in the decompression cycle. Our Raman scattering measurements confirmed that all the observations are reversible. 

\section{RESULTS AND DISCUSSION}

\subsection{High-pressure vibrational properties}

In a single molecule of TMGe, the germanium atom is tetrahedrally bonded to four methyl groups, as shown in Fig.~1. For ideally tetrahedral \emph{T}$_\emph{d}$ (-43\emph{m}) molecular symmetry, one bond distance C-H, one Ge-C, and one Ge-C-H angle would describe fully the molecular geometry. The methyl groups can be fully staggered with respect to the Ge-C bonds fully eclipsed or disordered [Fig.~1(b)]. The irreducible representation (of all the normal vibrational modes) is $\Gamma$=3$\emph{A}$$_1$+$\emph{A}$$_2$+4$\emph{E}$+4$\emph{F}$$_1$+7$\emph{F}$$_2$. According to the selection rule, the \emph{A}$_1$, \emph{E} and \emph{F}$_2$ vibrations are Raman active. The measured modes at
ambient conditions are summarized in Table I. Our data at ambient pressure are in good agreement with the results reported previously,\cite{Watari1978,Brown1960} except for the mode $\nu$$_5$. Due to weak and broad Raman signal inherent, the value of mode $\nu$$_5$ is not precise enough and thus it is omitted safely in the following discussion.

\begin{table*}
\caption{\label{tab:table1}Assignment of the observed Raman modes of TMGe, changes of Raman modes with pressures and the pressure coefficients of the corresponding frequencies of the Raman modes. Observed modes were taken at ambient pressure and room temperature in the liquid phase for all internal modes and appearing with compression in the proposed phases. Pressure coefficients $\textbf{d}\nu\textbf{/}\textbf{d}\emph{\textbf{P}}$ were obtained by linear fit of the Raman modes in four pressure regions, as indicated in Fig. 4.}
\begin{ruledtabular}
\begin{tabular}{c|c|ccccc|ccccc}
 \textbf{Modes of vibration} & \textbf{Type of vibration}\footnotemark[1] & \multicolumn{5}{c|}{\textbf{Obs. (cm$^{-1}$)}} & \multicolumn{5}{c}{$\textbf{d}\nu\textbf{/}\textbf{d}\emph{\textbf{P}}$\textbf{ (cm}$^{\textbf{-1}}$\textbf{/GPa)}}\\
 \cline{3-12}
 & &\textbf{ Liquid} & \textbf{I} & \textbf{II} & \textbf{III} &\textbf{ IV} & \textbf{Liquid} & \textbf{I} & \textbf{II} & \textbf{III} & \textbf{IV} \\ \hline
  &  &  & 120 &  &  & None &  &  4.9 & 4.9 & 5.4 &  \\
 $\nu_{1}$ & E (C-Ge-C skeletal deformation) & 181 &  &  &  &  & 39.4 & 12.4 &  &  & \\
  &  &  &  & 258 &  &  &  & 9.0 & 9.2 & 7.2 & \\
 $\nu_{2}$ & F$_{2} $ (C-Ge-C skeletal deformation) & 195 &  &  &  &  & 42.3 & 13.1 &  &  & \\
 $\nu_{3}$ & A$_{1} $ (C-Ge skeletal stretch) & 560 &  &  &  &  & 10.7 & 5.9 & 4.5 & 4.4 & 2.9 \\
 $\nu_{4}$ & F$_{2} $ (C-Ge skeletal stretch) & 599 &  &  &  &  & 7.1 & 5.5 & 4.1 & 4.6 & 2.6 \\
 $\nu_{5}$ & E (CH$_{3}$ rocking) & 820 &  &  &  &  &  &  &  &  & \\
  & F$_{2} $ (CH$_{3}$ rocking) &  &  &  &  &  &  &  &  &  & \\
  &  &  &  &  &  & 1220 &  &  &  &  & 1.9 \\
  &  &  & 1217 &  &  &  &  & 3.9 & 0.9 & 0.4 & 2.1 \\
 $\nu_{6}$ &  F$_{2} $ (CH$_{3}$ symmetrical deformation) & 1238 &  &  &  &  & -2.5 & 2.7 &  &  & \\
  &  & &  & 1240 & None &  &  &  & 3.7 &  & \\
 $\nu_{7}$ & A$_{1} $ (CH$_{3}$ symmetrical deformation) & 1249 &  &  &  &  &-7.3& 4.3&  &  & \\
 $\nu_{8}$ & E (CH$_{3}$ nonsymmetrical deformation) & 1400 &  &  &  &  &  &  &  &  & \\
 $\nu_{9}$ & F$_{2}$ (CH$_{3}$ nonsymmetrical deformation) & 1437 &  &  &  &  &  &  &  &  & \\
 &  &  &  &  &  & 3073 &  &  &  &  & 5.4 \\
 $\nu_{10}$ & E (CH$_{3}$ symmetrical stretch) & 2907 &  &  &  & & 20.8 & 11.9 & 7.5 & 7.4 & 4.7 \\
 & F$_{2}$ (CH$_{3}$ symmetrical stretch) &  &  &  &  &  &  &  &  &  &  \\
 $\nu_{11}$ & E (CH$_{3}$ nonsymmetrical stretch) & 2974 &   &   &  &  & 25.7 & 12.2 & 7.7 & 8.0 & 4.9 \\
 & F$_{2}$ (CH$_{3}$ nonsymmetrical stretch) &  &  &  &  &  &  &  &  &  &  \\
  &  &   &   &   &  & 3156 &  &   &   &   & 5.4 \\
\end{tabular}
\end{ruledtabular}
\footnotetext[1]{From Refs. [\onlinecite{shimizu1960,Overend1960}].}
\end{table*}

Vibrational spectroscopy is critical for characterizing the high-pressure behaviors of low-Z molecular compounds. Raman vibrational spectra of TMGe were collected from ambient pressure to high pressures up to 30.2 GPa, and the selected spectra are shown in Fig.~2. Clearly, the Raman spectra could be divided into four regions based on the molecular nature of the complex: the C-Ge-C skeletal deformation region (100-500 cm$^{-1}$), the C-Ge skeletal stretch region (500-800 cm$^{-1}$), the CH$_3$ symmetrical deformation region (1100-1300 cm$^{-1}$), and the CH$_3$ symmetrical and nonsymmetrical stretch region (2800-3300 cm$^{-1}$). With increasing pressure, all of the measured peaks shift to higher frequencies, and become weak at pressure up to 30.2 GPa. Some of peaks nearly vanish except for the $\nu$$_8$ = 1400 cm$^{-1}$ and $\nu$$_9$ = 1437 cm$^{-1}$ (the CH$_3$ nonsymmetrical deformation modes) which are influenced by the strong peak of diamond at 1332 cm$^{-1}$. This suggests that several pressure-induced phase
transformations take place in TMGe.

The $\nu$$_1$ and $\nu$$_2$ modes in the C-Ge-C skeletal deformation region are very close to each other at ambient pressure [Fig. 3(a)], and thus are difficult to be identified. At the onset of compression, the intensities of the $\nu$$_1$ and $\nu$$_2$ modes exhibit reversal changes, yet recover soon, which is evidence for the exchange the symmetry assignment of of the $\nu$$_1$ and $\nu$$_2$ modes as a result of Fermi resonance.\cite{Lin2008} Upon further compression up to 1.4 GPa, two new modes emerge, a lattice vibrational mode at low frequency region of 100-150 cm$^{-1}$ and a mode ($\nu$$^\prime$$_6$) slightly locating below the mode $\nu$$_6$. The emergence of new vibrational modes, especially the lattice mode, can be identified to the liquid-solid state transition with the application of the external pressure. Furthermore, the sharp peak of the lattice mode suggests the new phase with higher ordered structure, which is also found in other compounds.\cite{Xie2009} With increasing pressure to 3.0 GPa, the Raman spectra change dramatically, indicating substantial changes in the crystal and/or molecular structures. As shown in Fig.~2, the peak of the lattice mode becomes sharper and its intensity increases at around 3.0 GPa.
The most prominent change in the C-Ge-C skeletal deformation region is the mergence of $\nu$$_1$ and $\nu$$_2$ modes. In addition, the $\nu$$_7$ mode in the CH$_3$ symmetrical deformation region disappears at 3.0 GPa. Compressing continually to 5.4 GPa, the lattice mode $\nu$$_6$, which has sharper peak at the pressure range of 1.4 to 3.0 GPa, broadens and weakens. Another mode in the CH$_3$ symmetrical deformation region also disappears. With continuous compression, there is no obvious change in the number of peaks in all the four regions. These Raman modes only weaken and broaden with increasing pressure. This implies that the compound has become compact at this pressure range.

\begin{figure}[tbp]
\vspace{-0.5cm}
\includegraphics[width=\columnwidth]{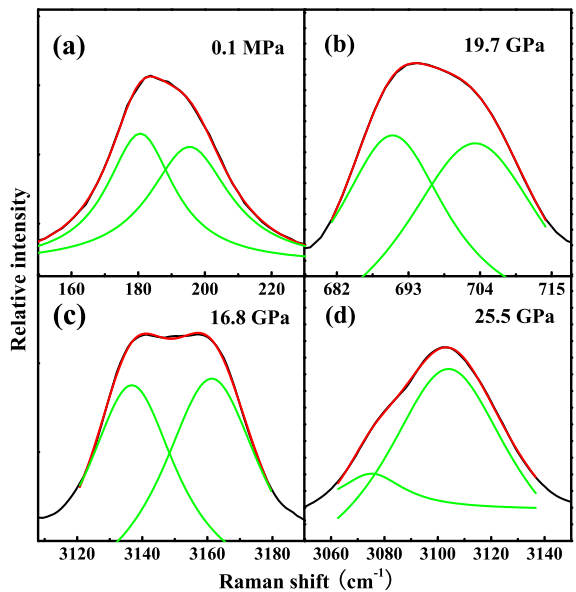}
\vspace{-1cm}
 \label{xijie} \caption{(Color online) Comparison of observed and fitted Raman spectra of TMGe with respect to overlapped modes of (a) $\nu$$_1$ and $\nu$$_2$ at ambient conditions, (b) $\nu$$_4$ with a shoulder at 19.7 GPa, (c) $\nu$$_1$$_0$ splitting with pressure up to 16.8 GPa, (d) $\nu$$_9$ accompanied to appearing of shoulder peak at 25.5 GPa. Black lines indicate observed Raman spectra; red lines through black lines indicate sum curve of the fit; the individual bands of the fit are represented by green lines in the lower part. }
\vspace{-0.2cm}
\end{figure}

Upon further compression to 16.8 - 20.3 GPa, rich Raman features are observed. As seen in Fig.~2, it is difficult to resolve the lattice mode in the C-Ge-C skeletal deformation region, and a new internal mode appears just below the mode $\nu$$^\prime$$_6$. A weak shoulder peak with a higher frequency of 691 cm$^{-1}$ is observed simultaneously as adjacent mode of $\nu$$_4$, yet disappears at 20.8 GPa. The mode of $\nu$$_1$$_1$ splits into two peaks. Proceeding
with compression, no change has been observed from all the modes except for the doubly degenerate mode of $\nu$$_1$$_0$, which arises distinctly a shoulder peak at 25.5 GPa with the frequency of 3073 cm$^{-1}$. It is difficult to identify whether the shoulder peak has already been existed at 16.8-20.3 GPa. In the whole compressed process, the development of several vibrational modes is exhibited by overlapping form in Raman signals. For this reason, Figs. 3(b), 3(c), and 3(d) show the overlapping modes separate by fitting the experimental data at various pressures.

\begin{figure}[tbp]
\includegraphics[width=\columnwidth]{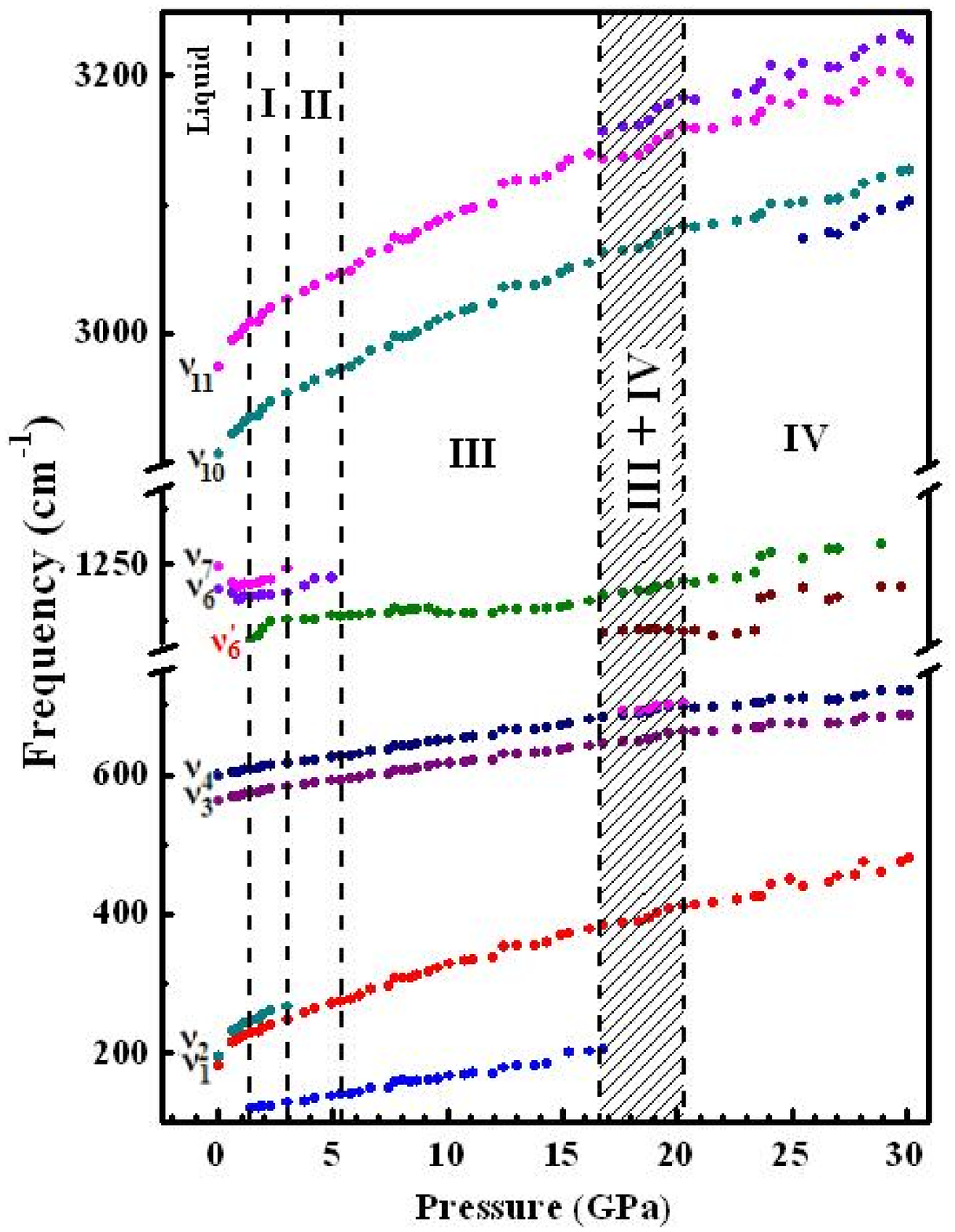}
\vspace{-0.5cm}
 \label{phase} \caption{(Color online) Pressure dependence of the frequencies of TMGe for the observed modes in all
regions at room temperature. The vertical dashed lines at near 1.4, 3.0, 5.4 and 16.8-20.3 GPa indicate the proposed phase boundaries.}
\vspace{-0.2cm}
\end{figure}

Vibrational frequencies provide information of the high-pressure behavior of TMGe. To show the possible phase transitions upon compression, the pressure dependences of Raman modes are depicted in Fig.~4. Clearly, there are several distinct pressure regions in Fig.~4 labeled by Liquid (below 1.4 GPa), phase I (1.4 - 3.0 GPa), phase II (3.0 - 5.4 GPa), phase III (5.4 - 16.8 GPa), and phase IV (above 20.3 GPa). In the liquid phase, the $\nu$$_6$ and
$\nu$$_7$ modes show a softening behavior, which is a typical character of rotational mode of CH$_3$ group,\cite{Adams1988,Rush1966,Adams1993} however, they exhibit blue-shift in the first solid phase (phase I). This softening only exists in the liquid phase and the related modes disappear at around 5 GPa, which is different from the case of TMS under pressure.\cite{qin2011} The softening of CH$_3$ groups of TMS remains until 9 GPa and the related modes do not disappear in the whole pressure region of 30 GPa. In phase I, the mode $\nu$$^\prime$$_6$ of TMGe also undergoes blue-shift, but its d$\nu$/d\emph{P} is larger than those of $\nu$$_6$ and $\nu$$_7$
modes, manifesting that new high-pressure structure is prone to be compressed. The pressure effect could be also observed in phases II and III in which a CH$_3$ rocking mode disappears and the C-Ge-C skeletal deformation modes merge. This indicates that CH$_3$ groups inside molecule are partly locked in the positions under pressure and the main skeleton of molecules deform due to the increased intra- and intermolecular interaction. In phase III, further compression restricts fully the movement from the deformation of CH$_3$ groups. Compression up to 16.8 - 20.3 GPa, the number of the Raman modes increases in the high-pressure phase IV, which indicates a new phase with lower symmetry.

Phase transformations are further identified by the changes of pressure coefficients of the Raman modes. The fitted pressure coefficients (d$\nu$/d\emph{P} (cm$^{-1}$/ GPa)) of the monitored peaks obtained by linear regression are listed in Table I. In general, most of the pressure coefficients of the stretch modes decrease noticeably with increasing pressure. From the liquid phase to phase I, the value of d$\nu$/d\emph{P} of the modes $\nu$$_6$ and $\nu$$_7$ changes from negative to positive, indicating the rotational motion of the CH$_3$ groups is compelled to be frozen in positions.\cite{Lin2008} Interestingly, there is an unusual case that high-pressure phase III possesses unexpected higher compressibility than that of phase II. This provides the powerful evidence of phase transition, although the number of Raman peaks nearly does not change in the both phases. Compared to the pressure coefficients of phase II and III, the phase IV shows a mutation in the most Raman modes, which suggests that new crystal structure would be made up with a closer-packing of atoms.

Structural transformation at different temperatures and pressures has been a critical issue to explore the feasibility of metallic hydrides. Compared to TMS under pressure, TMGe exhibits rich phase transitions at low pressures. Especially for the CH$_3$ groups, no rotational mode in intrinsic spherically shaped molecules of TMGe at ambient conditions is assigned, yet it exhibits softening vibration related to the rotation of CH$_3$ group(s).\cite{Adams1988,Rush1966,Adams1993} Although the external pressure makes CH$_3$ groups of TMGe locked in the positions and restricts the mobility of the hydrogen atoms, it is uncertain whether all hydrogen atoms are built up in a network structure by means of the closest packing,\cite{Ashcroft52,Chen53,Martinez54} which has significant implications for metallic hydrogen under pressure. Unfortunately, no direct evidence is found to illustrate the metallization of TMGe ($i.e.$, visible darkening of sample in DAC) under pressure up to 30 GPa. Very recently, it is reported that silane may undergo partial decomposition with compressed above 50 GPa, which hinders the search for the stable metallic hydrides. The high-pressure behaviors of TMGe, especially its stability and rich phase, manifest itself as a candidate of hydrogen-rich material for achieving metallization at high pressures.

\begin{figure}[tbp]
\vspace{-0.5cm}
\includegraphics[width=\columnwidth]{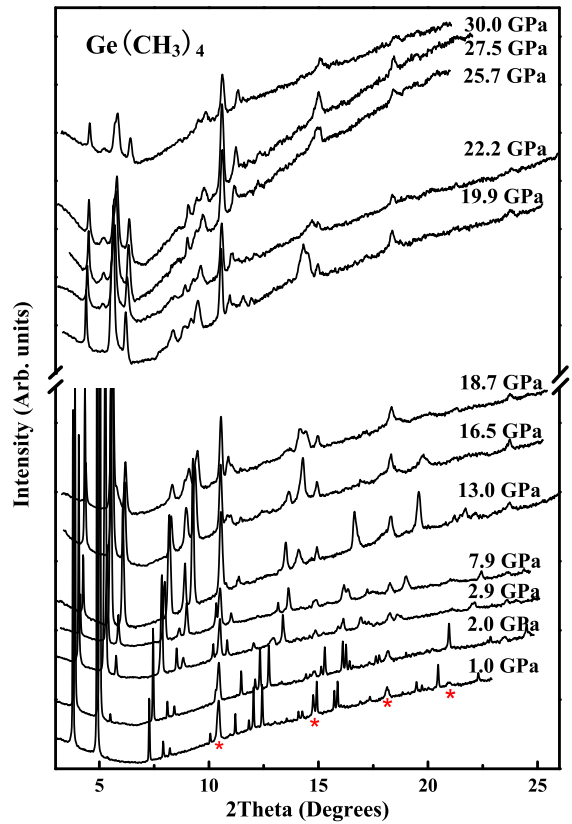}
\vspace{-1cm}
 \label{xrd} \caption{(Color online) Synchrotron X-ray ($\lambda=0.4067~\AA$) diffraction patterns of TMGe during the pressurization from ambient conditions to 30.0 GPa. Red asterisk marked to illustrate the signals from the gasket material tungsten.}
\vspace{-0.2cm}
\end{figure}

\subsection{Determination of high-pressure phases}

The pressure dependence of the scattering profile provides evidence for several phase transitions. Figure 5 shows the X-ray diffraction patterns of TMGe at pressures up to 30.0 GPa. All the Bragg peaks shift to larger angles, showing the shrinkage of the TMGe lattice. Upon compression to 2.9 GPa, there are several changes in the XRD patterns. The shape, intensity and width of the peaks are distinct from low-pressure patterns in the region above 10$^{\textordmasculine }$, and new peaks appear, which is consistent with the Raman results of phase transition at 3.0 GPa. In this pressure range, the changes of XRD patterns show the process of crystallization of TMGe from the liquid to solid state. With continuous compression to 7.9 GPa, obvious changes in the relative intensity of peaks are observed, which suggests the TMGe undergoes phase transition. This phase corresponds to the phase III observed in Raman spectroscopy above 5.4 GPa. Further compressed to 16.5 GPa, the XRD patterns again exhibit the changes in the number, intensity, and sharp of peaks until 19.9 GPa, coinciding with phase IV from the Raman data. From 20.0 to 30.0 GPa, the XRD patterns keep stable with weak and broadening peaks. In the whole region of compression, several phase transitions determined by the XRD patterns are in accordance with those observed by Raman spectroscopy.

To investigate the crystal structure of each phase, the diffraction patterns obtained at selected pressures were refined using Le Bail method with GSAS software.\cite{Larson1994} The phase I was fitted as the space group of \emph{Pa}-3, and the measured and fitted data are shown in Fig. 6(a). The obtained space group of \emph{Pa}-3 is consistent with the theoretical calculation,\cite{Wolf2010} in which the \emph{Pa}-3 phase was predicted by global lattice-energy minimizations using force-field methods. At 1.0 GPa, the TMGe has the lattice parameter of \emph{a} = 10.5968(4)~\AA, which is similar to that of TMS. At 0.58 GPa and 296 K, TMS has a lattice parameter of \emph{a} = 10.7328~\AA~[\onlinecite{Gajda2008}]. In \emph{Pa}-3 phase, the molecules are situated on three-fold axes, and thus CH$_3$ groups form a distorted cubic close packing, which is relatively rare among organic homomolecular crystals.\cite{Belsky1995}

\begin{figure}[tbp]
\vspace{0cm}
\includegraphics[width=\columnwidth]{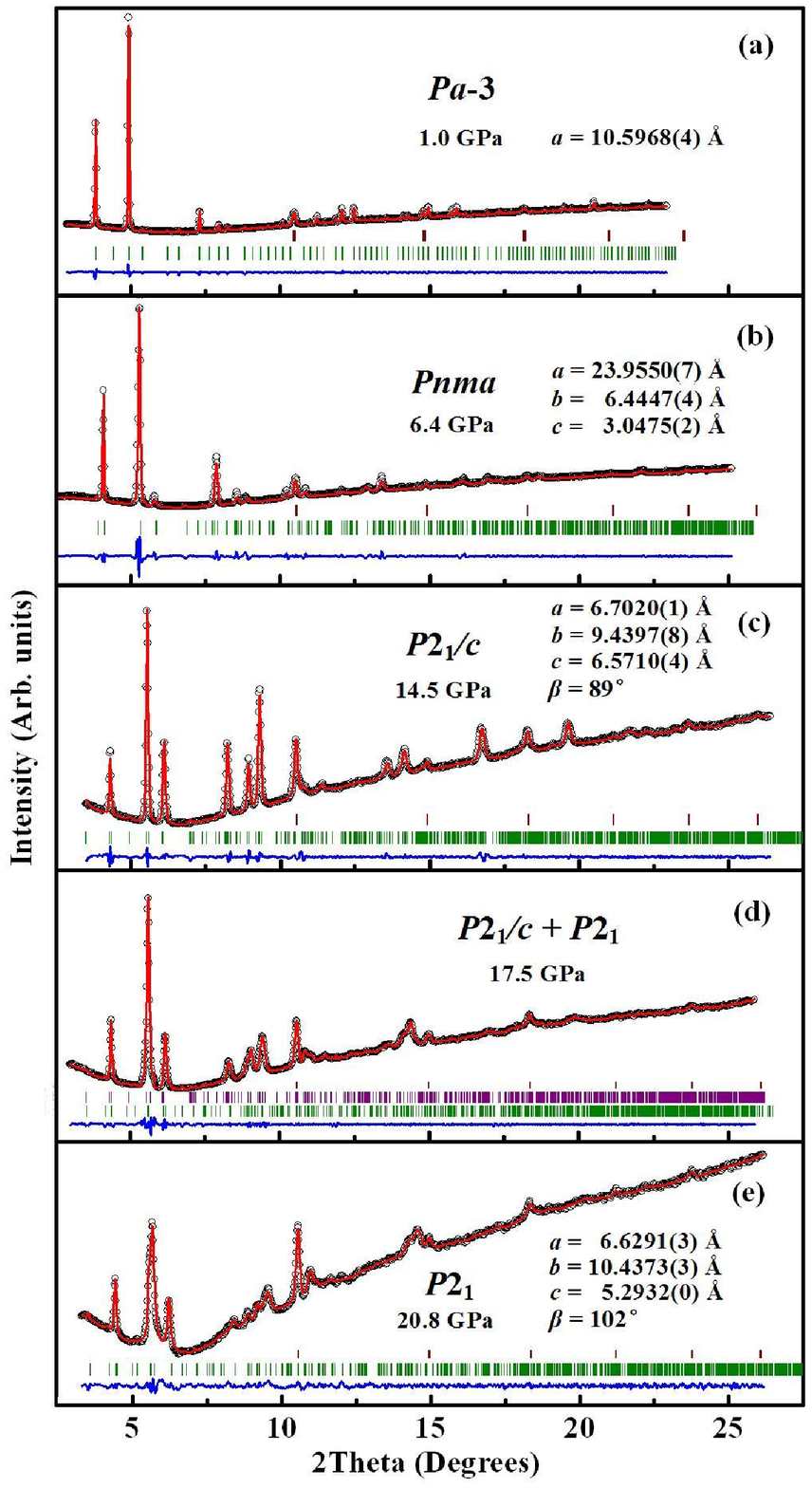}
\vspace{-0.5cm} \label{nihe} \caption{(Color online) X-ray powder diffraction patterns of liquid TMGe at pressures of 1.0 (a), 6.4 (b), 14.5 (c), 17.5 (d) and 20.8 (e) GPa. The refined lattice parameters for the corresponding space groups are given respectively. The open circles represent the measured intensities and the red lines the results of profile refinements by the best LeBail-fit with each space group. The positions of the Bragg reflections are marked by vertical lines and the difference profiles are shown at the bottoms (blue lines). The \emph{R} values are \emph{R}$_\emph{p}$=0.7\%, \emph{R}$_\emph{wp}$=1.0\% for the fitting at 1.0 GPa, \emph{R}$_\emph{p}$=0.8\%,
\emph{R}$_\emph{wp}$=1.6\% at 6.4 GPa, \emph{R}$_\emph{p}$=0.4\%, \emph{R}$_\emph{wp}$=0.7\% at 14.5 GPa, \emph{R}$_\emph{p}$=0.3\%, \emph{R}$_\emph{wp}$=0.6\% at 17.5 GPa, and \emph{R}$_\emph{p}$=0.3\%, \emph{R}$_\emph{wp}$=0.5\% at 20.8 GPa.}
\vspace{-0.2cm}
\end{figure}

There is no information of crystal structure available at higher pressures for TMGe. The possible crystal structures of the unknown phases were analyzed with the program Dicvol06 [\onlinecite{Louer2007}] and Peakfit v4. For phase II, 10 peaks were resolved at 2.9 GPa and indexed mainly to the orthorhombic system. As minority, the cubic and tetragonal systems were ruled out due to unreasonable figures of merit (M,F) and/or volume of the cell. There are several space groups allowing for the orthorhombic phase, such as \emph{Pnma}, \emph{P}2$_1$2$_1$2$_1$, \emph{Pbca}, \emph{Ama}2, \emph{Cmcm}, and \emph{Pmn}2$_1$ from the predicted crystal structures of TMGe.\cite{Wolf2010} Among them, \emph{Pnma} is a strong candidate for phase II because it shows better fit to the diffraction profile at 2.9 GPa. Additionally, the space group \emph{Pnma} (Z=4) of TMGe was suggested as the second best structure energetically\cite{Wolf2010} at ambient pressure. Figure 6(b) shows the fitted results of the phase II with \emph{Pnma} space group at 6.4 GPa with lattice parameters of \emph{a} = 23.9550(7)~\AA, \emph{b} = 6.4447(4)~\AA, and \emph{c} = 3.0475(2)~\AA.

\begin{figure}[tbp]
\includegraphics[width=\columnwidth]{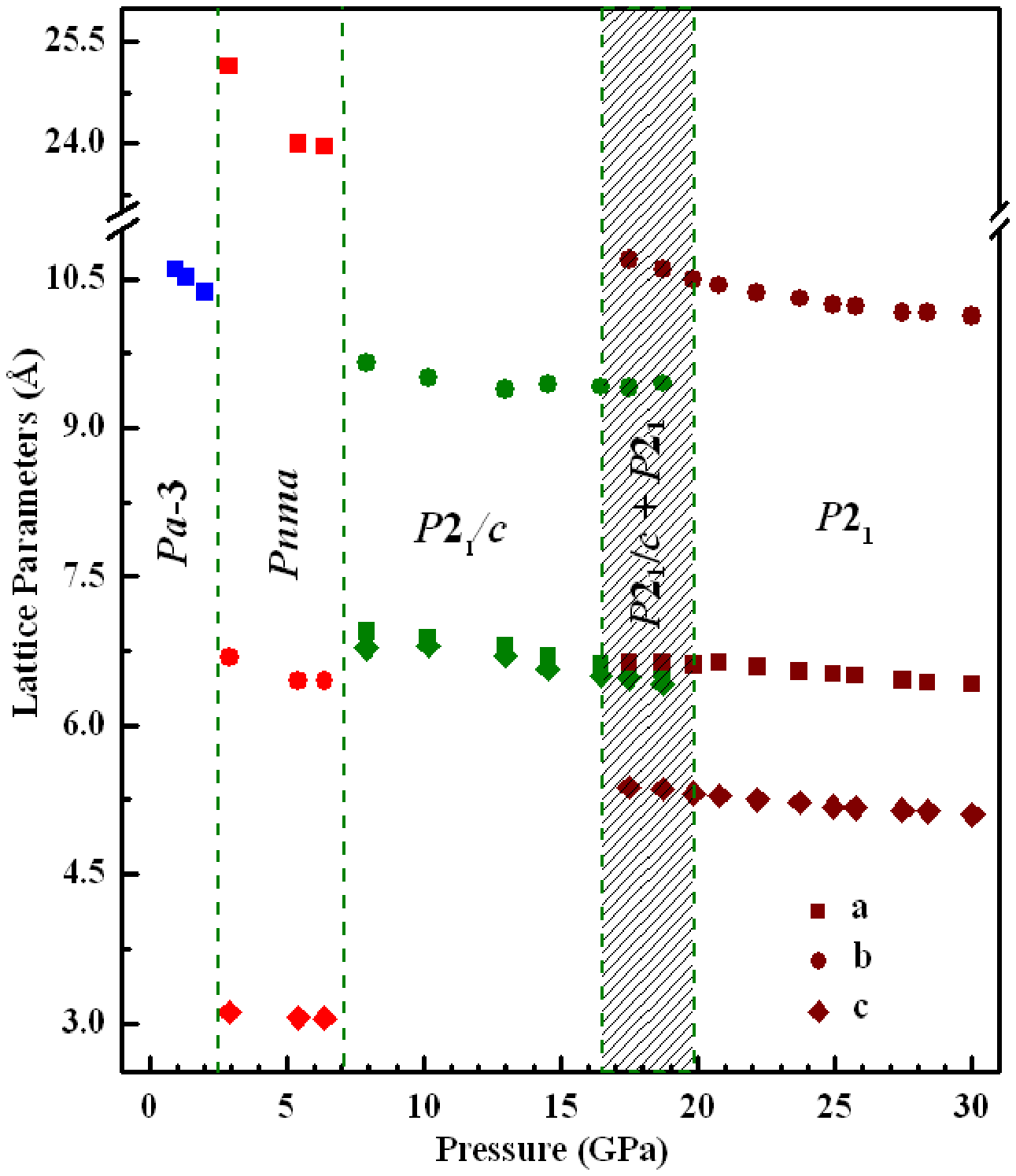}
\vspace{-0.5cm} \label{canshu} \caption{(Color online) Pressure dependence of the lattice parameters corresponding space groups for each phase, the vertical dashed lines denote the phase boundaries.}
\vspace{-0.2cm}
\end{figure}

For phase III of TMGe, other 13 peaks from the XRD pattern at 7.9 GPa were resolved and indexed mainly to the monoclinic system and one orthorhombic system. Among the above plausible space group in orthorhombic system, the predicted volume per formula unit is 94.62~\AA$^3$ assuming Z = 4. However, this volume is illogical because it leads to considerable compressibility of 17~\AA$^3$/GPa per formula unit when comparing with the result of V =
117.62~\AA$^3$ per formula unit at 6.4 GPa. The related high-pressure studies on the sister compounds,
C(Si(CH$_3$)$_3$)$_4$ [\onlinecite{Dinnebier2000}], Si(C(CH$_3$)$_3$)$_1$(Si(CH$_3$)$_3$)$_3$ [\onlinecite{Wunschel2003}], and Si(C(CH$_3$)$_3$)$_2$(Si(CH$_3$)$_3$)$_2$ [\onlinecite{Wunschel2003}] have
found their small compressibilities below 6 GPa. For monoclinic system, it is difficult to determine space group for this new phase, whereas \emph{P}2$_1$/\emph{c} is a candidate because that tetrahalides of the IVa groups elements, MX$_4$ (M = Si, Ge, Sn; X = Cl, Br) with halogen atoms have a comparable size to a methyl
group crystallize in \emph{P}2$_1$/\emph{c}, Z = 4 [\onlinecite{Zakharov1986,Reuter2000,Reuter2001,Merzand2002,Kohler2005,Wolf2009}]. Additionally, \emph{P}2$_1$/\emph{c} space group with Z = 4 was also predicted as the third best structure of TMGe energetically and appeared repeatedly with increasing energy of crystal structure.\cite{Wolf2010} Considering the Raman results that CH$_3$ groups are locked in positions and the whole groups move like one atom, the \emph{P}2$_1$/\emph{c} space group would be the most reasonable solution to the structure of TMGe at 7.9 GPa. Figure 6(c) shows the result to fit the patterns of TMGe at 14.5 GPa by the space group of \emph{P}2$_1$/\emph{c}. Compared to GeCl$_4$ (\emph{a} =
9.6903~\AA, \emph{b} = 6.4508~\AA, \emph{c}= 9.7740~\AA, and $\beta$ = 103.075$^{\textordmasculine }$ ) and GeBr$_4$ (\emph{a} = 10.1832~\AA, \emph{b} = 6.7791~\AA, \emph{c}= 10.2922~\AA, and $\beta$ = 102.543$^{\textordmasculine }$) at low temperature,\cite{Merzand2002,Wolf2009} the lattice parameters of \emph{a} = 6.7020(1)~\AA, \emph{b} = 9.4397(8)~\AA, \emph{c} = 6.5710(4)~\AA, and $\beta$ = 89.036$^{\textordmasculine }$ are debatable. A plausible cause for the abnormity is non-hydrostatic situation of crystallized TMGe at pressures, which leads to the lattice distortion.

\begin{figure}[tbp]
\includegraphics[width=\columnwidth]{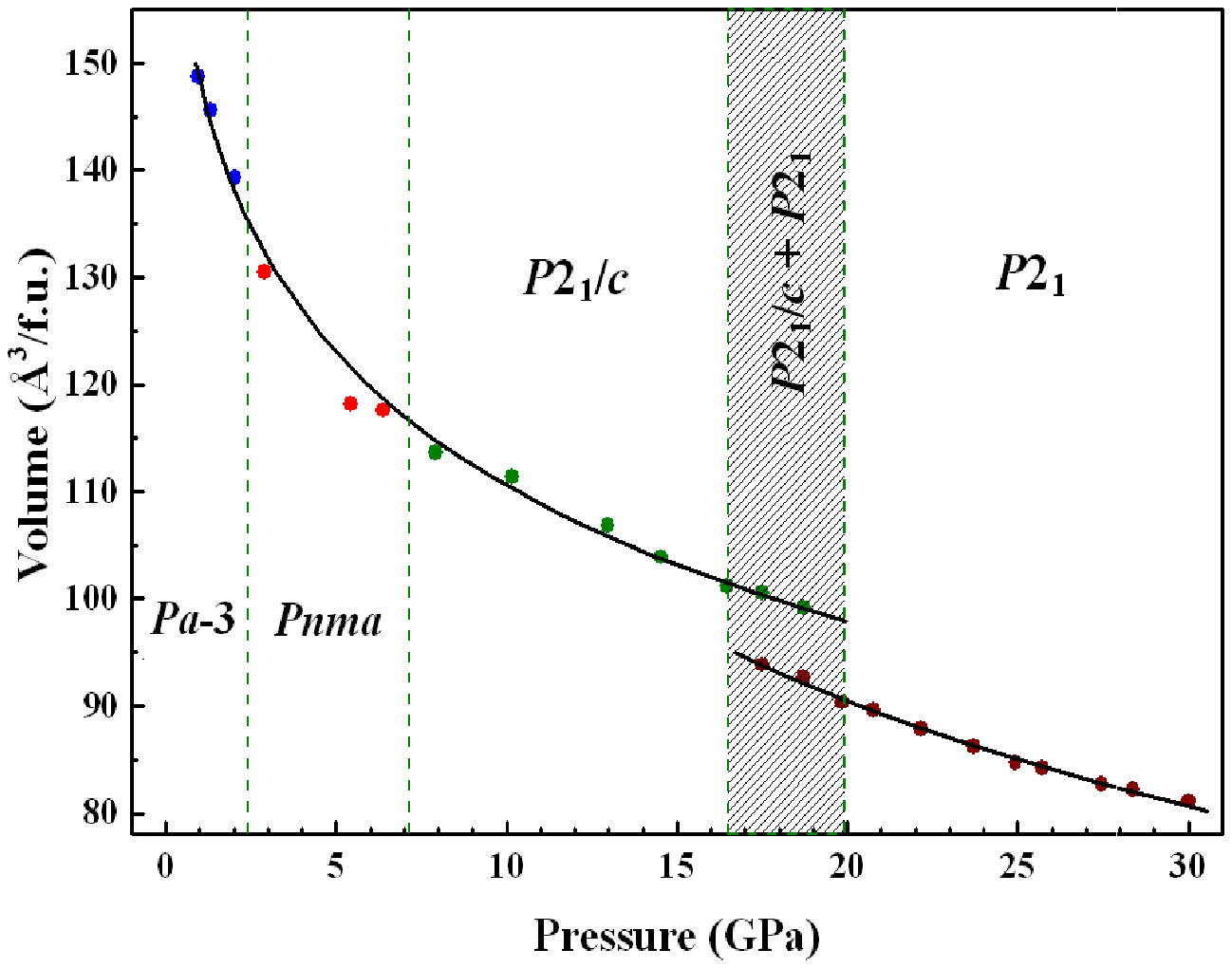}
\vspace{-0.5cm}
\label{canshu} \caption{(Color online) Volume per formula unit change of TMGe with pressure. The solid lines demonstrate the fitting data of phases to the Birch-Murnaghan equation of state and the vertical dashed lines denote the phase boundaries.}
\vspace{-0.2cm}
\end{figure}

For phase IV, the diffraction patterns yield two orthorhombic systems and four monoclinic systems at 19.9 GPa. According to the Raman results in phase IV, the number of Raman bands greatly increase, indicating the phase IV with lower symmetry. Along with the unreasonable values of V = 637.90 ~\AA$^3$ and V = 617.87 ~\AA$^3$ given in the lattice parameters of orthorhombic systems, the orthorhombic systems have been ruled out. For monoclinic systems, only one of the four lattice parameters with space group \emph{P}2$_1$, Z = 4 in all of the monoclinic system was available
for corresponding systematic extinction rule. The lattice parameters are \emph{a} = 6.7127~\AA, \emph{b} = 10.5933~\AA, \emph{c} = 5.4200~\AA, and $\beta$ = 99.980$^{\textordmasculine }$. Figure 6(e) shows the measured and fitted patterns of TMGe at 20.8 GPa. The problem still remains regarding whether space groups \emph{P}2 and \emph{Pm}
in monoclinic system also conform to the XRD patterns in phase IV due to no constraint of systematic extinction rule. Indeed, the refining results of \emph{P}2 and \emph{Pm }in monoclinic system also bring out the similar results as the case by \emph{P}2$_1$ at phase IV. For the transient phase, only space group \emph{P}2$_1$ and \emph{P}2$_1$/\emph{c} are selected to fit the pattern at 17.5 GPa, as showed in Fig. 6(d).

\subsection{Equations of states}

To reveal the compressibility of each phase, the lattice parameters and volume per formula unit were computed by fitting the patterns at selected pressures from the GSAS software.\cite{Larson1994} The changes of cell parameters are illustrated in Fig.~7, in which we only selected \emph{P}2$_1$ space group as a possible situation for phase IV, and Fig.~8 plots the pressure dependence of the volume per formula unit as a function of pressure for each phase. The volume/pressure relationship represents the equation of state (EoS), which can be described analytically by series expansions of Eulerian finite strain such as the Birch-Murnaghan equation of state (BM3 EoS)\cite{Birch1947} defined as
\begin{equation}
\emph{P}=3\emph{K}_0\emph{f}_E(1+2\emph{f}_E)^\frac{5}{2}[1+\frac{3}{2}(\emph{K}_0^\prime-4)\emph{f}_E] ~~\nonumber ,
\end{equation}
where $\emph{f}_E=[(\frac{\emph{V}_0}{\emph{V}})^\frac{2}{3}-1]$, \emph{V}$_0$ is the volume at ambient pressure, \emph{V} is the volume at pressure \emph{P} given in GPa, \emph{K}$_0$ is the bulk modulus at 0 GPa, and \emph{K}$_0$$^\prime$ is the first pressure derivative of \emph{K}$_0$. The solid lines in Fig. 8 represent the
fitted Birch-Murnaghan EoS, which yields more accurate parameters for highly compressible materials. 

The bulk moduli of phase \emph{Pa}-3, \emph{Pnma}, and \emph{P}2$_1$/\emph{c} are 2.19 $\pm$ 0.08 GPa with \emph{K}$_0$$^\prime$= 15.00 $\pm$ 0.13, $\emph{V}_0=175.02\pm0.58 ~{\AA}^3$ and phase \emph{P}2$_1$ is 9.47 $\pm$ 0.65 GPa with \emph{K}$_0$$^\prime$= 4.00$\pm$ 0.01, $\emph{V}_0=167.26\pm3.07 ~\AA^3$, respectively. For phase \emph{Pa}-3, \emph{Pnma}, and \emph{P}2$_1$/\emph{c}, the results are coincident with the analogous compound of C(Si(CH$_3$)$_3$)$_4$ [\onlinecite{Dinnebier2000}], indicating a soft feature. The increasing bulk moduli infer an enhancement of bond strength during phase transitions and indicate the intrinsic higher compressibility. It is worth to mention that phase \emph{P}2$_1$ has a relatively large bulk modulus compared with the high-pressure phases of TMS. This suggests that crystal structure of phase \emph{P}2$_1$ has been entirely transformed and that the layered network would be possible in view that the homologous compound, TMS, had started to form layers along (011) lattice plane in the \emph{Pnma} phase at low temperature.\cite{Wolf2010} The layered crystal structure for hydrogen atoms has been suggested to be an essential  metallic state in hydrogen-bearing compounds.\cite{Feng2006,chen2008,Gao2008,Tsejs2007,li2010}
Furthermore, the bulk modulus in phase \emph{P}2$_1$ is remarkable strength because silane (SiH$_4$) has gotten the bulk modulus of 7.89 GPa upon compressed to 39 GPa [\onlinecite{Degtyareva2007}] and SiH$_4$-H$_2$ complex
has also achieved the value of 6.87 GPa with pressure up to 35 GPa [\onlinecite{Strobel2009}], whereas methane (CH$_4$) could get the same value only by compressing to 13 GPa [\onlinecite{Nakahata1999}]. Recently, high-pressure studies \cite{tim,zhong} on hydrogen-rich germanium compounds GeH$_{4}$-H$_{2}$ revealed very rich vibrational dynamics, intermolecular interactions, structural, electronic, and potential superconducting properties. 

So far, there is little information on such a magnitude of hydrogen-bearing compound TMGe. Measurements of electronic transport properties are expected to be performed in order to examine whether TMGe would undergo metallization and eventually become a superconductor at higher pressures. It should be noticed that a recent electronic transport study on molecular hydrogen revealed a significant resistance drop at 260-270 GPa [\onlinecite{Eremets2011}]. However, two independent measurements \cite{alex,zha} indicate that metallic hydrogen has not been reached yet even at 360 GPa. The softening of some Raman modes on CH$_3$ groups and their sudden disappearance in Ge(CH$_3$)$_4$ indicate that this compound might be ideal for metallization and even high-temperature superconductivity at modest static pressure for laboratory capability.

\section{CONCLUSIONS}

We performed Raman measurements of TMGe at room temperature  and at pressures up to 30 GPa. Our results revealed the phase transitions at 1.4, 3.0, 5.4 and 20.3 GPa from the mode frequency shifts with pressure. We found that phase transitions of TMGe are more sensitive to pressure than those of tetramethylsilane. These transitions were suggested to result from the changes in the inter- and intra-molecular bonding of this material. Further work using synchrotron X-ray radiation revealed three phase transitions at 2.9, 7.9, and 19.9 GPa with similar results from Raman measurements. The space groups for the high-pressure phases were determined to be \emph{Pa}-3 for phase I, \emph{Pnma} for phase II, \emph{P}2$_1$/\emph{c} for phase III, and \emph{P}2$_1$ for phase IV. The equations of states were obtained up to 30 GPa. Such structural information may be helpful in exploring possible superconductivity in hydrogen-bearing compounds at high pressures.

\begin{acknowledgments}
We are grateful to Zhiqiang Chen for his technical support during the experiment. X.J.C. acknowledges EFree, an Energy Frontier Research Center funded by DOE-BES under grant number DE-SC0001057. This work in China was supported by the Cultivation Fund of the Key Scientific and Technical Innovation Project Ministry of Education of China (No.708070), the Shenzhen Basic Research Grant (No. JC201105190880A), the National Natural Science Foundation of China (No. 11274335) and Guangdong Natural Science Foundation (No. S2012040007929), and the Fundamental Research Funds for the Central Universities SCUT (No.2012zz0078). Use of the National Synchrotron Light Source, Brookhaven National Laboratory, was supported by the U.S. Department of Energy, Office of Science, Office of Basic Energy Sciences, under Contract No. DE-AC02-98CH10886.
\end{acknowledgments}

\end{document}